\documentclass[12pt]{JHEP3}
\usepackage{amsfonts}
\usepackage{epsfig}
\def\ol#1{\overline{#1}}
\def\Dsl{\hbox{/\kern-.6700em\it D}} 
\def\dsl{\hbox{/\kern-.5300em$\partial$}}
\def\veps{\varepsilon}

\def\eqa{\begin{eqnarray}}
\def\eeqa{\end{eqnarray}}
\def\eeq{\end{equation}}
\def\nn{\nonumber}

\def\veps{\varepsilon}
\def\pref#1{(\ref{#1})}

\def\roughly#1{\mathrel{\raise.3ex\hbox{$#1$\kern-.95em
\lower1ex\hbox{$\sim$}}}}

\def\exd{{\rm d}}

\def\ssubsubsection#1{\vspace{3mm} \noindent \textbf{#1} \\ \vspace{-3mm} \\ \noindent}

\def\be{\begin{equation}}
\def\ee{\end{equation}}
\def\bea{\begin{eqnarray}}
\def\eea{\end{eqnarray}}

\newcommand{\ba}{\begin{eqnarray}}
\newcommand{\ea}{\end{eqnarray}}
\newcommand{\n}{\nonumber\\}

\title{D-Brane Chemistry}

\author{C.P. Burgess,$^1$ N.E. Grandi,$^2$ F. Quevedo$^2$ and R. Rabad\'an$^3$
\\

$^1$ Physics Department, McGill University,  3600 University Street,
 Montr\'eal, Qu\'ebec, Canada, H3A 2T8.\\

$^2$ Centre for Mathematical Sciences, DAMTP,
               University of Cambridge,\\
               Cambridge CB3 0WA UK.\\

$^3$ Theory Division CERN, CH-1211 Gen\`eve 23, Switzerland
}

\abstract{We study several different kinds of bound states built
from D-branes and orientifolds. These states are to atoms what
branonium -- the bound state of a brane and its anti-brane -- is
to positronium, inasmuch as they typically involve a light brane
bound to a much heavier object with conserved charges which forbid
the system's decay. We find the fully relativistic motion of a
probe \Dpp-brane in the presence of source \Dp-branes is
integrable by quadratures. Keplerian conic sections are obtained
for special choices for $p$ and $p'$ and the systems are shown to be
equivalent to non-relativistic systems. Their quantum behaviour is also
equivalent to the corresponding non-relativistic limit.
In particular  the $p=6$, $p'=0$ case is equivalent to a
non-relativistic dyon in a magnetic monopole background, with trajectories
in the surface of a cone.
 We also show that the motion of
probe branes about $D6$-branes in IIA theory is equivalent to the
motion of the corresponding probes in the uplift to M theory in 11
dimensions, for which there are no $D6$-branes but their fields
are replaced by a particular Taub-NUT geometry. We further
discuss the interactions of D-branes and orientifold planes having
the same dimension. This system behaves at large distances as a
brane-brane system but at shorter distances it does not have the
tachyon instability.}


\preprint{McGill-03/18, DAMTP-2003-89}

\def\be{\begin{equation}}
\def\ee{\end{equation}}
\def\bea{\begin{eqnarray}}
\def\eea{\end{eqnarray}}

\def\Dp{$Dp$}
\def\Dpp{$Dp'$}

\begin{document}


\section{Introduction}
We consider the motion of \Dpp-branes moving in the presence of
sources, which we take to be either \Dp-branes having $p > p'$ or
$Op$ (orientifold) planes having $p = p'$. These systems can
contain the brane analogs of atoms, inasmuch as they can form
stable orbits which are stable against annihilation or decay by
virtue of the conserved charges which their constituents carry.
(This represents an important difference between these systems and
the branonium systems --- {\it i.e.} a \Dp-brane and
$\overline{D}p$-antibrane --- studied in \cite{branonium1}.)

We find that the \Dp-\Dpp\ systems experience an attractive
interaction precisely for those choices of $p - p'$ which
correspond to the existence of a tachyon in the open-string
spectrum, which presumably therefore indicates the instability
toward the formation of a supersymmetric bound state built from
the \Dp-\Dpp\ pair. We compute the classical trajectories (and
some quantum properties) for the motion of the \Dpp-brane in the
probe-brane limit. For some choices the resulting orbits are
simply the usual conic sections of the non-relativistic Kepler
problem, even though they apply to the fully relativistic motion
of the probe brane. Our study of these systems is motivated by the
remarkable properties of the related branonium systems, for which
the fully relativistic motion of a probe antibrane through the
fields sourced by a stack of source branes may be completely
reduced to quadratures. We find this property to be shared by the
broader class of systems consisting of \Dpp-branes orbiting
\Dp-brane sources.

 \FIGURE{\centering \hspace*{0in}\vspace*{.2in}\epsfxsize=3in
    \epsffile{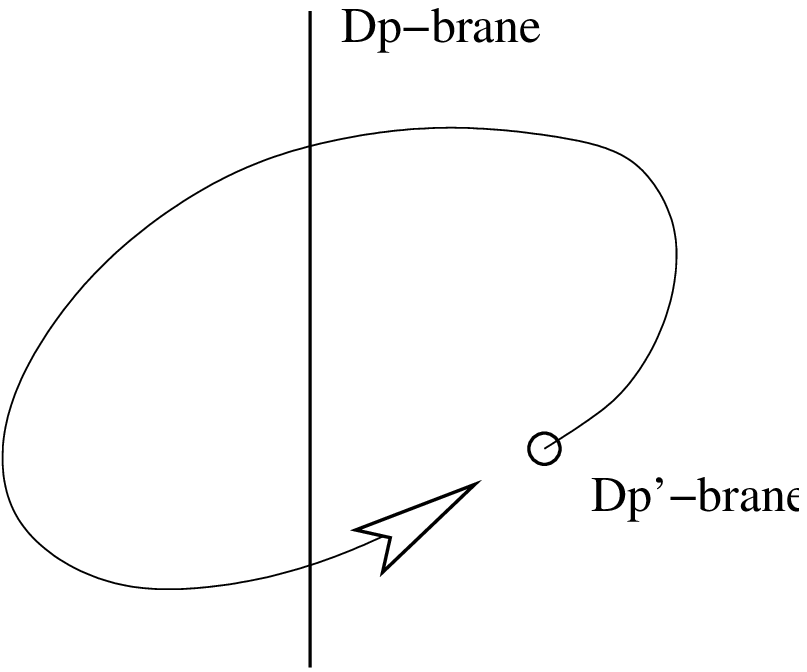}\hspace*{0.4in}
    \caption{\small A $Dp'$-brane orbits a higher dimensional $Dp$-brane. }
    \label{Dp-Dp-prime}}

Our presentation is organized as follows. Section 2, immediately
following, sets up and solves the equations of motion for the
probe \Dpp-brane system in the presence of $N$ source \Dp-branes.
Section 3 then examines how this motion looks for the special
situation of IIA theory where the source branes are $D6$ branes.
In this case the theory may be uplifted to M-theory in 11
dimensions with the $D6$ branes disappearing, but with the fields
they source being replaced by an extended Taub-NUT geometry. We
show in this section that the motion of probe branes about
$D6$-branes in IIA theory is equivalent to the motion of
equivalent probes in the Taub-NUT space. Section 4 then examines
several features of the interaction potential between \Dp-branes
and $Op^\pm$ orientifold planes. We close with our conclusions in
section 5.
\section{Brane Atoms: $D{p'}$-branes in $Dp$-background}
In this section we generalize the results of \cite{branonium1} to
the situation where the dimension, ${p'}$, of the probe brane is
different from the dimension, $p$, of the source brane (see figure
\ref{Dp-Dp-prime}). We consider the higher-dimensional \Dp-brane
(or set of \Dp-branes) as the source of a background field
configuration within which the probe \Dpp-brane moves. This is a
good approximation when the mass of the higher-dimensional brane
(or set of branes) is bigger than the mass of the
lower-dimensional one. In particular, it is always the case when
the directions transverse to the lower-dimensional brane but
parallel to the higher-dimensional one are very large or
non-compact, and we henceforth assume this to be true for the
remainder of this section.


\subsection{Constituents}
We begin by reviewing the relevant properties of the background
fields and the probe branes.

\ssubsubsection{The Background}
We consider the \Dp-brane to be static, acting only as a source
for a supergravity background. This means that we ignore the
massive closed-string modes which could also have been sourced,
and so implies the validity of a low-energy expansion ({\it i.e.}
we neglect $\alpha'$ corrections). Since these modes become
relevant at short distances compared with the string scale, $r \ll
l_s$, at these distances the supergravity framework breaks down
and the system is better described within the open-string picture.


The background sourced by the higher-dimensional $Dp$-brane is
given by the following solution to the bosonic field equations of
ten-dimensional supergravity
\ba
    ds^2 &=& h^{-1/2}(-dt^2 + dx_1^2 + \cdots + dx_p^2 ) +
    h^{1/2}(dy_{p+1}+\cdots+dy_{9}) \n
    C_{0\cdots p}&=&1-\frac 1 h \qquad \hbox{and} \qquad
    e^\phi = h^{({3-p})/4}
    \label{background}
\ea
where $C_{(p+1)}$ is the $(p+1)$-dimensional Ramond-Ramond (RR)
form which couples to brane charge, and $\phi$ is the dilaton.
When $p < 7$ the function $h$ is given explicitly by
\be
    h = 1 + \frac k {r^{7-p}}
\ee
with $r^2 = y_{p+1}^2 + \cdots + y_9^2$ and $k = c_p \,g_s N \,l_s^{7-p}$,
being $c_p =(2  \sqrt \pi)^{5-p}\,\Gamma(\frac{7-p}2)$.


\ssubsubsection{Open-String Tachyons and Stability}
As mentioned above, we take the \Dpp-brane to be a probe which
moves in the above background, created by the higher-dimensional
\Dp-brane. Although this description works well for large brane
separations, it fails at distances of the order of $l_s$, where
there can be open string modes that become massless or can become
tachyonic. We now analyse the stability of the system from this
open-string point of view, which we shall see nicely complements
our later analysis of the equation of motion of the probe brane as
it moves through the background fields.

In the brane-antibrane case, an open-string tachyon appears once
the branes are separated by $r \sim l_s$, signalling the system's
instability towards brane-antibrane annihilation. A similar
tachyon develops for \Dpp-branes orbiting \Dp-branes if $|p'-p| <
4$. If $|p'-p| = 4$ the would-be tachyon mode is instead a
massless scalar mode and the static system is supersymmetric in
the sense that the brane system preserves $1/4$ of the bulk
supersymmetries. (This supersymmetry and flat direction is
reflected in the absence of a potential of interaction between the
branes when they are parallel and are not in relative motion.)
When $|p'-p| > 4$ the lowest open-string mode at short distances
is massive.

As we shall see, the existence of this tachyonic open-string mode
at short distances correlates with whether the one-loop brane
interaction potentials are attractive --- a result which further
generalizes to the more generic case of branes at angles or branes
with magnetic fluxes. We find the following generic rule:
attractive (repulsive) potentials appear when the lowest
open-string mode at short distances is tachyonic (massive). The
static (supersymmetric) cases correspond to the lowest mode being
massless. This correlation has a simple physical interpretation:
the existence of an open-string tachyon signals an instability
towards the formation of a bound state of the two branes through
the attractive interaction potential, leading to a lower-energy
({\it i.e.} tension) configuration, which is typically
supersymmetric and stable (and presumably represents the global
energy minimum). One would expect from this interpretation that
the two branes are attracted to one another in order to reach this
minimum, and this is what we shall find.


\ssubsubsection{The action}
In the background (\ref{background}), we introduce a probe
$Dp'$-brane, whose action is given by a Born-Infeld (BI) and
Wess-Zumino (WZ) term, according to
\ba
    S&=&-T_{p'}\int d^{p'+1}\xi \, e^{-\phi}
    \sqrt{|g_{AB}\partial_\alpha X^A\partial_\beta
    X^B|}-q\,T_{p'}\!\int\! C_{p'}
    \n&& \qquad\qquad\qquad\qquad\qquad \alpha,\beta=0,...,p'\;;
    \qquad\qquad A,B=0,...,9 \,,
    \label{ac}
\ea
where $X^A(\xi_\alpha)$ represents the embedding of the brane into
spacetime, and $|M_{\alpha\beta}|$ denotes the absolute value of
the determinant of any matrix $M_{\alpha\beta}$. In this
expression the world-volume gauge field
$F_{\alpha\beta}(\xi_\alpha)$ has been turned off, $T_{p'}$ is
the tension of the probe brane and $q=\pm 1$ the corresponding RR
charge.


After gauge-fixing the reparameterization invariance of the action
(\ref{ac}) using the so-called static gauge, $X^\alpha(\xi^\beta)
= \xi^\alpha$, we assume the brane moves rigidly ({\it i.e.}
$X^a = X^a(t)$) which means that we neglect any oscillatory
modes of the brane\footnote{These modes must be considered when
analysing the stability of the solution, as in
\cite{branonium1}.}.
The probe brane is also imagined to be oriented parallel to some
of the dimensions of the higher-dimensional source brane during
this rigid motion. Under these circumstances the probe brane moves
like a particle in the $9-p$ dimensions transverse to the source
brane. The BI part of the action then becomes
%
\be
    S_{BI} =
    -m \int dt \,h^{(
    {p-3})/4}\sqrt{\left| g_{\alpha\beta} + \delta_\alpha^0\delta_\beta^0
    \dot X_a \dot X^a \right|}
    \qquad\qquad
    a,b=p'+1,...,9 ,
\ee
where $g_{\alpha\beta} = h^{-1/2}\delta_{\alpha\beta}$. Since our
interest is in the case ${p'} < p$, we can assume the directions
parallel to the probe brane are toroidally compactified without
any need to modify the background field configurations. With this
understanding, the factor $m$  in this last equation is the `mass' of
the
\Dpp-brane $m = T_{p'}V_{p'}$ with $V_{p'}$ the
volume of these compact dimensions.


{}From here on we reserve the coordinates $y^n$ (with $n =
p+1,...,9$) for the dimensions transverse to the \Dp-brane, and
the coordinates $x^i$ (with $i = p'+1,...,p$) for those dimensions
transverse to the \Dpp-brane, but parallel to the \Dp-brane. With
this choice we have $\dot X_a \dot X^a=h^{-1/2}\dot x_i^2+
h^{1/2}\dot y_n^2$. Evaluating the determinant finally
allows the BI part of the action to take the general form
\ba
    S_{BI} &=& -m\int dt\, h^{(
    {p-{p'}} - 4)/4}\sqrt{1-\dot x_i^2 - h \, \dot y_n^2} \n &&
    \qquad\qquad\qquad\qquad\qquad i={p'}+1,...,p\;; \qquad\qquad
    n=p+1,...,9 \, .
    \label{accion1}
\ea


The WZ term can differ from zero in any of the following three
circumstances: ($i$) If $p = p'$; ($ii$) If the \Dp-brane carries
a lower-dimensional charge ({\it e.g.} a bound state of the
\Dp-brane with lower-dimensional branes); or ($iii$) If there are
several probe branes and a Myers-type of effect develops
\cite{m99}, {\it etc.} In this section we consider \Dp-branes
without magnetic flux and we take only a single probe. Since $p
\neq {p'}$ and the world-volume gauge fields on the probe brane
are set to zero, the Ramond-Ramond fields do not affect the
movement of the probe brane. Therefore, generically at long
distances the probe brane only feels the influence of gravity and
the dilaton field.


The only possible WZ interaction which can arise between the
\Dpp-brane and the background is a coupling to the dual, $\tilde
C_{p'}$, whose field strength, $\tilde F = \exd \tilde C_{p'}$ is
the Hodge dual of the field strength of the background
$(p+1)$-form, $F = \exd C_{p}$. For this coupling to be possible
$\tilde C_{p'}$ must be a $(p'+1)$-form, and so requires
$p+{p'}=6$. For an arbitrary external ${p'}$-form $\tilde C_{p'}$,
after gauge fixing and assuming rigid motion the WZ action becomes
\ba
    S_{WZ}=  -mq\int dt \left(\tilde C_{0\cdots {p'}} +
    \tilde C_{a 1\cdots {p'}}\,\dot X^a\right)
    \qquad\qquad a=p'+1,...,9 \,,
\ea
where $q=1$ for a brane and $q=-1$ for an antibrane.
Furthermore, assuming $\tilde
C_{{p'}} = \tilde C_{{p'}}(z)$ for some subset, $z$, of the
coordinates, the curvature must have nonzero components $\tilde
F_{01\cdots {{p'}}z}$ and $\tilde F_{a1\cdots p' z}$. Being the
Hodge dual of the $F_{0\cdots p}$ curvature sourced by the
background brane, it cannot have components in the $0$ direction,
and so $\tilde F_{01\cdots p' z}=0$, which in turn implies that
$\tilde C_{0\cdots {{p'}}}$ must vanish. Moreover, since the
$1\cdots {{p'}}$ directions of the probe are parallel to some of
the $1\cdots p$ directions of the background, the dual field
strength, $\tilde F_{a1\cdots {{p'}}z}$, cannot have components in
these directions. From this last statement we deduce that $\tilde
C_{a1\cdots p'} = 0$ if ${{p'}}\neq 0$. We conclude that a
nonvanishing WZ term is possible only when $p'=0$ and then $p=6$.

\subsection{Equations of motion}
We next examine the equations of motion for the probe brane, and
with the previous paragraph in mind we divide the discussion into
two cases, depending on whether or not there is a WZ contribution
to the probe-brane action. When discussing the various
possibilities it is worth keeping in mind that our choice of
background fields assumes $p < 7$ and that $D$-branes are all odd-
or all even-dimensional within any particular string theory, and
so we may take $p - p'$ to be even. It follows that the difference
$p - p'$ can take the possible values $0,2,4$ and 6, although we
exclude the case $p-p'=0$ here, because this is the case studied
in the branonium analysis \cite{branonium1}.



\ssubsubsection{Born Infeld Only}
We first assume ${{p'}} \neq 0$ or $p \neq 6$, so that the WZ term
does not appear in the probe-brane action. Since this excludes the
case $p-p'=6$, we need consider only $p - p' = 2$ or $4$ (with
this last possibility corresponding to the supersymmetric case if
the branes are at rest).


Rotational invariance in the $y_n$ space ensures the conservation
of angular momentum, $L_{mn} = y_m p_n-y_n p_m$, and so we can
describe the plane in which the motion takes place using polar
coordinates, $r,\varphi$. The action (\ref{accion1}) then becomes
\ba
    S_{BI} &=& -m\int dt\, h^{(
    {p-{p'}}-4)/4}\sqrt{1-\dot x_i^2 - h (\dot r^2 + r^2\dot \varphi^2)}
    \quad\;\;\;  i={p'}+1,...,p\;, \label{accion2}
\ea
and the canonical momenta are
\ba
    p_i &=& m\, \frac {h^{({p-{{p'}}}-4)/4}} {\sqrt{(\cdots)}} \dot
    x_i\equiv m \, u_i\n
    p_r &=& m \frac {h^{({p-{{p'}}})/4}}
    {\sqrt{(\cdots)}}\dot r \equiv
    m\, \rho \n
    p_\varphi &=& m  \frac {h^{({p-{{p'}}})/4}} {\sqrt{(\cdots)}} r^2 \dot
    \varphi \equiv m\, \ell\,,
\ea
where $\sqrt{(\cdots)}$ stands for the square root in
(\ref{accion2}), and the right-hand-most equalities define the
variables $u_i$, $\rho$ and $\ell$.

Given these canonical momenta we can build the conserved
hamiltonian, which turns out to be
\be E = m \sqrt{ h^{({p-{{p'}}}-4)/2} + u_i^2 + \frac 1 h
\left( \rho^2 + \frac{\ell^2}{r^2} \right)} \equiv m
\,\varepsilon \, .
\label{Hamil}
\ee
Since this is a monotonically increasing function of the canonical
momenta, it is bounded from below by the following
`potential':
\be
V \equiv \varepsilon(\rho=u_i=\ell=0)= h^{({p-{{p'}}}-4)/4}
\ee
which, as expected, is a constant only in the supersymmetric case
$p-{{p'}}=4$\footnote{In order to see the vanishing of $V$ in the
case $p = {{p'}}$ one must also include the Wess-Zumino term,
which gives an extra contribution to the potential of the form $-
q h^{-1}$, where $q=1$ for a brane and $q = -1$ for the antibrane.
}. Notice that this potential is attractive (and so may have bound
states) if $p - p' < 4$  and repulsive if $p - p' > 4$ (as will be
shown in the next section), conditions which we have seen
also define when an open-string tachyon exists. At long distances, 
$r \gg l_s$, the potential goes as $V \sim 1
+ \frac14 \, (p - p' -4)\, k /r^{7-p}$.


Translation invariance within the dimensions parallel to the
\Dp-brane and rotational invariance in the dimensions transverse
to it ensure that the momenta $u_i$ and $\ell$ are conserved.
Their conservation allows the construction of an effective
potential for the radial motion in the usual way, with the result
\be
    V_{\rm eff}\equiv \varepsilon(\rho=0) = \sqrt{
    h^{({p-{{p'}}}-4)/2}+ u_i^2 + \frac {\ell^2} {hr^2} } \, .
\ee
See figures 2, 3 and 4 for plots of the shape of these potentials
given various choices for $p$ and $p'$. This potential shows
qualitatively when bound orbits can occur, since they do if the
effective potential ever passes below its asymptotic value at $r
\to \infty$. Circular orbits occur at the special radii which are
minima of $V_{\rm eff}$.

To determine when closed orbits exist we therefore explore the shape of
the potential near the origin and at infinity. In the
limit\footnote{Of course, if $r$ is taken all of the way to $l_s$
these expressions all receive string corrections, which become
important. Our interest is in $r$ not quite this small, to sketch
the shape of the potential.} $r\to 0$ we can replace $h$ by
$k/r^{7-p}$, obtaining
\be
    V_{\rm eff} = \sqrt{ b \, r^{ (p-7) {({p-{{p'}}}-4)/2}
     } + u_i^2 + \frac {\ell^2 r^{5-p}}k } \,, \qquad b =
    k^{(p-{{p'}}-4)/2} \,,
\ee
and since we assume $p<7$, the sign of the exponent in the first
term is given by the sign of $4 + p' - p$ ({\it i.e.} it is
positive when the interaction is attractive and negative when it
is repulsive). We find in this way the following limits:

\begin{itemize}
\item
For $p=6$ the last term ensures that the potential diverges at the
origin.
\item
For $p=5$ the last term tends to a constant for small $r$. For $p'
= 3$ the potential converges to $\sqrt{u_i^2+ \ell^2/m}$ and if
$p' = 1$ (the BPS case) it instead approaches $\sqrt{1+ u_i^2+
\ell^2/m}$.
\item
For $p<5$, the last term vanishes and at the origin the potential
converges to $\sqrt{u_i^2}$ if ${p-{{p'}}=2}$,
or to $\sqrt{1+ u_i^2}$ in the supersymmetric
case $p-p'=4$ ($i.e.~p=4,{{p'}}=0$).
\end{itemize}


\begin{figure}[h]
  \epsfxsize=3in \epsfysize=2in \label{p-pp}
  \begin{center}
  \begin{picture}(230,170)(-10,-10)
  \put(220,0){$r$} \put(110,42){{$~^{BPS~case}$}}
  \put(110,20){{$~^{p-p'<4}$}} \put(0,150){$V_{\rm eff}$}
  \put(190,150){{${p=6}$}} \put(0,0){\leavevmode\epsfbox{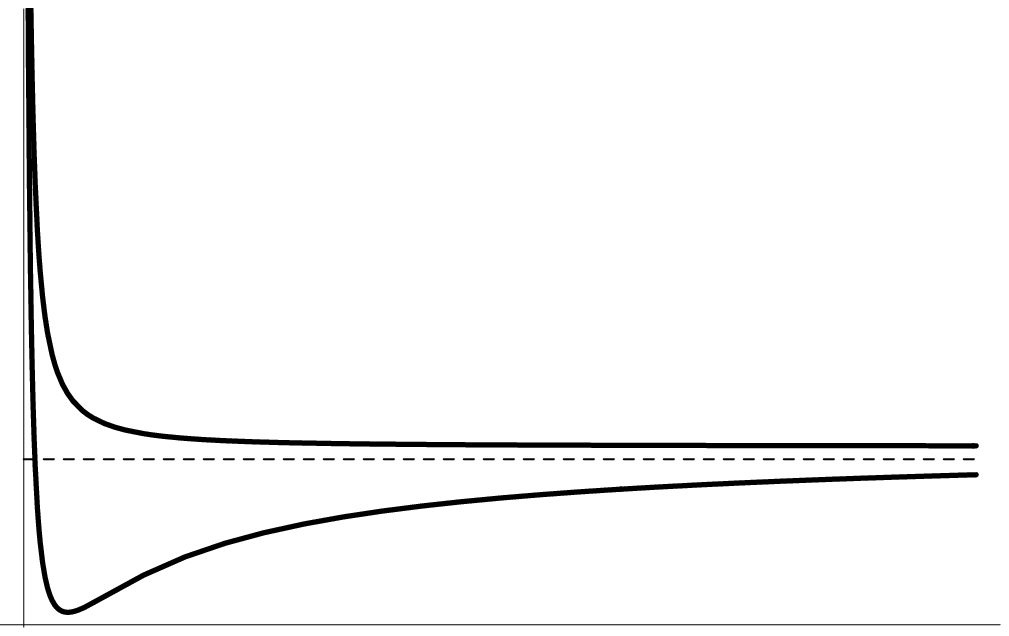}}
  \end{picture}
  \caption{\it Effective potential for the case $p=6$, in the BPS ($p'=
  2$) and non-BPS ($p'=4$) cases.}
  \end{center}
\end{figure}



On the other hand, in the large-$r$ limit, we instead find
\ba
    V_{\rm eff} &=& \sqrt{
    1+u_i^2+ \frac{(p-p'-4) k}{2 \, r^{7-p}} + \frac
    {\ell^2} {r^2} } \,,
\ea
and we see that the potential asymptotes to the constant
$\sqrt{1+u_i^2}$ when $r\to\infty$. This limit corresponds to the
expected dispersion relation for motion parallel with the
\Dp-brane, for which $E = \sqrt{m^2 + p_i^2}$ and so $\varepsilon
= \sqrt{1 + u_i^2}$. How this limit is approached depends on $p$
and $p'$, in the following ways.
%
%
\begin{itemize}
\item
In the supersymmetric case, $p-{{p'}}=4$ the limit is reached from
above, regardless of the value of $p$.
%
%
%
%
\item
If $p=6$, the non-supersymmetric case is $p' = 4$ and so the third
term dominates. This implies the potential approaches its limit
from below.
\item
For $p=5$ it is the choice $p' = 3$ which is not supersymmetric.
In this case the limiting value is reached from below or from
above depending on the sign of $\ell^2 + (p-p'-4)k/2 = \ell^2 -
k$.
\item
If $p<5$, the last term always dominates and the potential
approaches its limit from above.
\end{itemize}

\begin{figure}[h]
   \epsfxsize=3in \epsfysize=2in \label{g3}
   \begin{center}
   \begin{picture}(230,170)(-10,-10)
   \put(220,0){$r$} \put(7,40){{$~^{k(4-p+p')/2>\ell^2}$}}
   \put(51,100){{$~^{BPS~case}$}} \put(0,150){$V_{\rm eff}$}
   \put(7,68){{$~^{k(4-p+p')/2<\ell^2}$}} \put(190,150){{$p=5$}}
   \put(0,0){\leavevmode\epsfbox{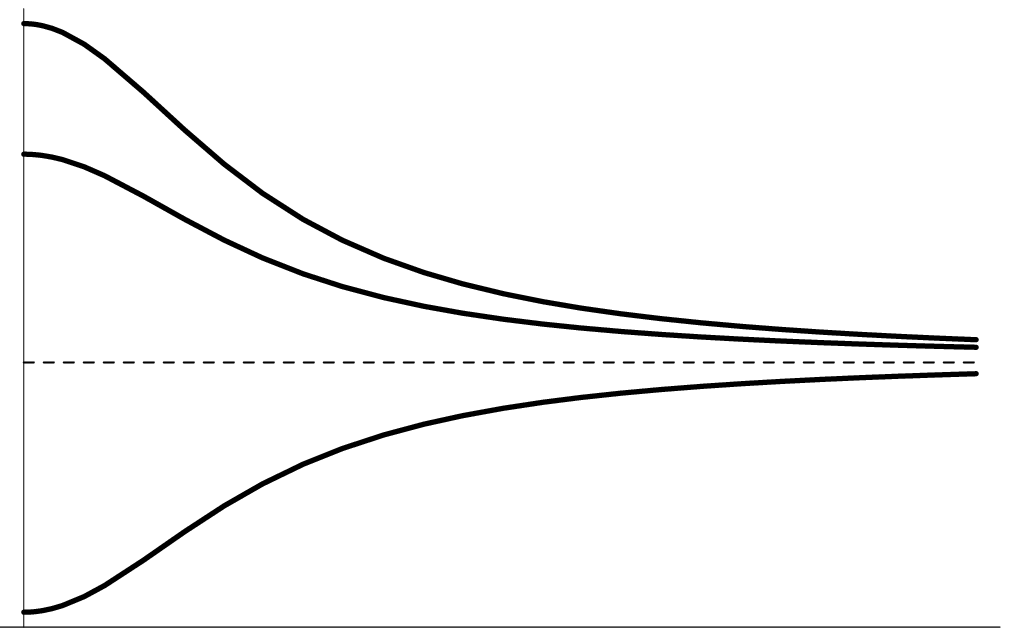}}
   \end{picture}
  \caption{\it Effective potential for the $p=5$ case, depending on the
   value of the angular momentum.}
   \end{center}
\end{figure}



Combining the behaviour for large and small $r$ the following
picture emerges.

\begin{itemize}
\item
If $p=6$ and ${{p'}}=4$ the $V_{\rm eff}$ reaches a minimum away
from $r=0$, and so bound orbits exist which are bounded away from
$r=0$. In the supersymmetric case where $p = 6$ and $p' = 2$ no
such minimum or bound states arise.

\item
If $p = 5$ and $p' = 1$ (the BPS case) there is no bound state. If
$p' = 3$ the existence of a minimum depends on the angular
momentum, and there is a bound orbit if $\ell^2 < (4 - p + p') k/2
= k$.

\item
If $p < 5$ the potential has a minimum at the origin. For the
supersymmetric case ($p = 4, p' = 0$) $V$ has the same value at
this minimum as it does at infinity. Otherwise $V$ is smaller at
$r = 0$ than at infinity. In either case classically localized
orbits exist.
\end{itemize}

\begin{figure}[h]
   \epsfxsize=3in \epsfysize=2in \label{g2}
   \begin{center}
   \begin{picture}(230,170)(-10,-10)
   \put(220,0){$r$} \put(58,138){{$~^{BPS~case}$}}
   \put(42,60){{$~^{p-p'<4}$}} \put(0,150){$V_{\rm eff}$}
   \put(190,150){{$p<5$}} \put(0,0){\leavevmode\epsfbox{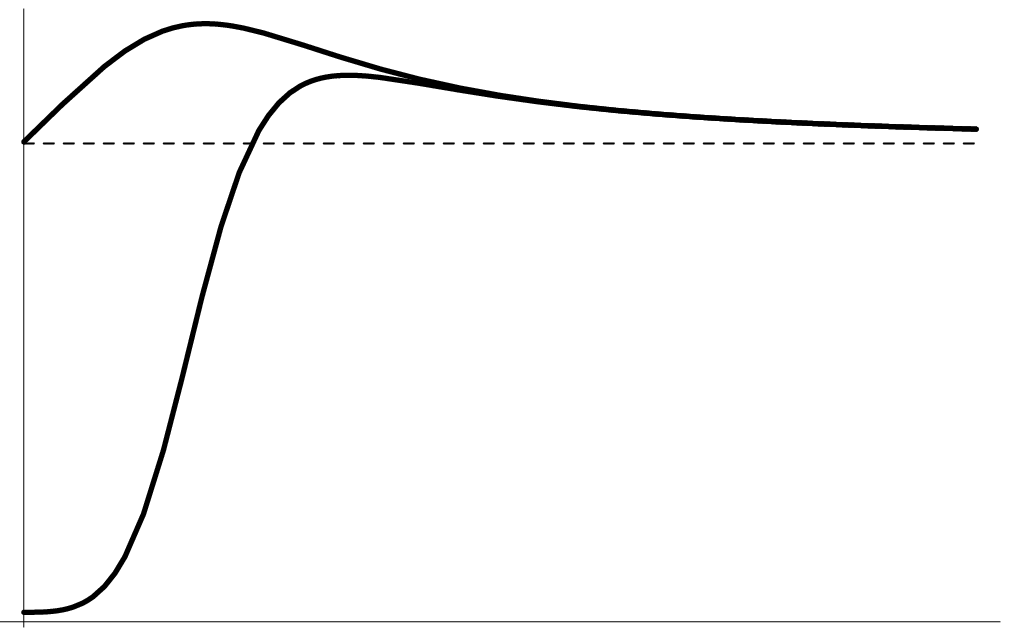}}
   \end{picture}
  \caption{\it Form of the effective potential when
  $p<5$.}
  \end{center}
\end{figure}



In terms of the conserved quantities $u_i$, $\ell$ and $\veps$,
the equations of motion reduce to first-order conditions
\ba
    \dot x_i &=& \frac{u_i}\varepsilon
    \n
    \dot r &=&  \frac{\rho}{\veps  h} =
    \frac 1 {\varepsilon h}
    \sqrt{({\varepsilon^2}-u_i^2)h
    -  h^{({p-{{p'}}}-2)/2} -\frac {\ell^2}{r^2}}
    \n
    \dot \varphi &=&  \frac{\ell}{\veps  h r^2}
 \,.
\ea
These equations of motion are easily integrated to obtain the
orbits
\be
    \varphi-\varphi_0 = \int  \frac {\ell \, dx} {\sqrt{({\varepsilon^2}-u_i^2)h
    -  h^{({p-{{p'}}}-2)/2} - {\ell^2}{x^2}}} \,,
    \label{orbit}
\ee
where the integration variable is $x=1/r$. Since $p-p'=2$ or 4,
the power of $h$ appearing here is $(p-p'-2)/2 = 0$ or $1$.
Consequently, for $p<6$ the quantity inside the square root is a
polynomial of degree $7-p$ in the variable $x$. It is quadratic
for both $p=5$ and $p=6$.
%

When the argument of the square root is quadratic in $x$ the
integral of \pref{orbit} reduces to
\be
    \varphi-\varphi_0 = \int\! \frac {dx} {\sqrt{-Ax^2 + Bx+ C}} \,
\label{inte} \ee
for appropriate constants $A, B$ and $C$ (which are listed in
table \ref{ABC}), then the integral may be performed explicitly to
give the standard conic sections of the Kepler problem.
\begin{table}
\begin{center}
\begin{tabular}{|cc|ccc|}
\hline
  $p$ & $p'$ & $A$ & $B$ & $C$ \\ \hline
  5 & 1 & $1 + k(1+u_i^2 -\veps^2)/\ell^2$ & 0 & $(\veps^2 - u_i^2 - 1)/\ell^2$ \\
  5 & 3 & $1 + k(u_i^2 - \veps^2)/\ell^2$ & 0 & $(\veps^2 - u_i^2 - 1)/\ell^2$ \\
  6 & 2 & 1 & $k(\veps^2 - u_i^2 -1)/\ell^2$ & $(\veps^2 - u_i^2 - 1)/\ell^2$ \\
  6 & 4 & 1 & $k(\veps^2 - u_i^2)/\ell^2$ & $(\veps^2 - u_i^2 - 1)/\ell^2$ \\ \hline
\end{tabular}
  \caption{The constants appearing in the orbit integral when
  trajectories are conic sections.}
  \label{ABC}
\end{center}
\end{table}
For instance, if $B \ne 0$ (which from the table requires $p = 6$)
the bound orbits are ellipses,
\be
    r = \frac{a(1-e^2)}{1 + e \cos(\sqrt{A} \varphi )} \,,
\ee
where we define
$a = - \, {B}/({2 C)}$ and $e^2 =  1 + 4AC/B^2$. The condition for
a bound orbit is $e < 1$, {\it i.e.} $C<0$, and this can be seen
from the table to require energies below the asymptotic value of
the potential $V_{\rm eff}$: $\varepsilon < \sqrt{1+u_i^2}$. Since
$V_{\rm eff} > 1 + u_i^2$ for all $r$ when $p' = 2$, we see that
Keplerian bound orbits arise only in the case $p = 6$ and $p'=4$.

For the case $p=5$, the orbits are also given by the integral
(\ref{inte}), but with $B=0$. If $A<0$ and $C<0$, we instead have
bound (non closed) orbits only if $p' = 3$ and $\ell^2 < k$, with
the form
%
%
\be
 r = \sqrt{\frac {A} C}\frac1{ \cosh(\sqrt{-A}\varphi)} \,.
\ee
The parameter $C$ can be negative only in the non-BPS case $p'=3$.
The sign of $A$ depends on $\ell$ and will be negative when
$\ell^2<(\varepsilon^2-u_i^2)k$, which is always the case if the
potential is attractive $\ell^2<k$ and the energy is less than the
value of the potential at infinity $\varepsilon < \sqrt{1+u_i^2}$.

The integral, eq.~\pref{orbit}, is similarly integrable in terms
of elliptic functions in the cases $p=1,3,4$.

\ssubsubsection{The Wess-Zumino term}
%
%
We now return to the special case for which a probe-brane coupling
with a WZ term is possible: $p=6$ and $p'=0$. In this case the
curvature sourced by the background brane has components
$F_{0\cdots 6r}=-k/(r^2h^2)$, and so the dual is $\tilde
F_{\theta\varphi} = -k \sin\theta$, where $r$, $\theta$ and
$\varphi$ are spherical polar coordinates in the three dimensions
transverse to the $D$6-brane. The gauge potential which produces
this dual field strength then is $\tilde
A_\varphi=k(\cos\theta-1)$, which corresponds to that of a
magnetic monopole and we have fixed the constant of integration by
demanding that the Dirac string is along the negative $z$ axis.
This leads to the WZ term
\be
    S_{WZ}= -qkm\int dt \,(\cos\theta-1) \,\dot \varphi
    \,.
\ee
This action arises more prosaically in the low-energy description
of spin waves in ferromagnets \cite{CBPhysRep}. The complete
action then reads
\ba
    S &=& -m  \int\! dt\left[ h^{1/2 }\sqrt{1-\dot x_i^2
    - h (\dot r^2 + r^2(\dot \theta^2 + \sin^2\!\theta\dot \varphi^2))}
    + qk(\cos\theta-1)\dot \varphi \right] .\;\;
\ea
%

In the absence of the WZ term the $D$0-brane motion would be confined to
a plane by conservation of angular momentum, $\vec L \propto \vec
y \times \dot{\vec y}$, transverse to the $D$6-brane. Choosing the
$z$ axis to be in the $\vec L$ direction then would allow the
$D$0-brane motion to be restricted to the equatorial plane,
$\theta = \pi/2$. The same is not true once the Wess-Zumino term
is included, because this term is not separately invariant under
3-dimensional rotations and gauge transformations, but is
invariant under a combination of both. Therefore, $\vec L$ is no
longer conserved and the motion need no longer be restricted to
lie on a plane. The conserved quantity in this case is a
generalized angular momentum, given by
\be
 \vec J = \frac {h^{3/2}} {\sqrt{(\cdots)}}\,\vec y \times \dot {\vec
 y} + qk\, \check y\,,
\ee
where $\check y = \vec y/|\vec y|$ is the unit vector in the $\vec
y$ direction. Choosing $\vec J$ to define the $z$ axis, the
implications of $\vec J$ conservation are most easily seen by
using the observation that $\vec J \cdot \check y = J \cos\theta =
qk$ is a constant of motion, and so the trajectories are
restricted to be at a fixed value of $\theta$, given by $\cos
\theta = qk/J$. That is, $\vec y$ precesses around a cone of
opening angle $\theta$ whose axis is in the $\vec J$ direction.
Notice that this cone becomes the plane $\theta = \pi/2$, as
expected, if we put $q = 0$.

For motion on this cone the action becomes
\ba
    S &=& -m  \int\! dt\left[ h^{1/2 }\sqrt{1-\dot x_i^2
    - h (\dot r^2 + r^2\sin^2\!\theta\dot \varphi^2)}
    + qk(\cos\theta-1)\dot \varphi \right] .\;\;
\ea
where $\cos\theta$ and $\sin\theta$ should be regarded as
constants which are determined from the relation $\cos\theta =
qk/J$. The canonical momenta are easily derived
\ba
    u_i &=&  \frac {h^{1/2}} {\sqrt{(\cdots)}}\, \dot x_i\n
    \rho &=&  \frac {h^{3/2}}
    {\sqrt{(\cdots)}}\,\dot r
 \n
    \ell &=&   \frac {h^{3/2}} {\sqrt{(\cdots)}} \,r^2\sin^2\theta\, \dot
    \varphi -qk(\cos\theta-1)\,,
\ea
and, as before, $u_i$ and $\ell$ are conserved because the action
is independent of $x^i$ and $\varphi$. The conserved hamiltonian
is
\be
 \varepsilon
 =  \sqrt{ h + u_i^2 + \frac 1 h
 \left( \rho^2 + \frac 1 {r^2}\left(
 \frac{\ell-qk(\cos\theta-1)}{\sin\theta }\right)^2 \right)}
 \, ,
\label{Hamil6}
\ee
and the potential derived from this hamiltonian is given by
\be
 V \equiv \varepsilon(\rho=u_i=\ell=0) = \sqrt{ h + \frac 1{hr^2}
 \left( \frac{k(\cos\theta-1)}{\sin\theta\, }\right)^2}
 \ =\ \sqrt{h+\frac{k^2}{hr^2}\left(\frac{J-qk}{J+qk}\right)}\,,
\ee
which is repulsive, as stated in the previous section.

The equations of motion obtained from the above hamiltonian are
\ba
 \dot x_i &=&\frac{ u_i}\varepsilon\n
 \dot r &=& \frac \rho{h\varepsilon}
 =\frac 1 {h\varepsilon}
 \sqrt{
 (\varepsilon^2-u_i^2)h-h^2 -\frac{\hat\ell^2}{r^2}
 }\n
 \dot \varphi &=& \frac 1 {h\varepsilon}\,
 \frac{\hat\ell}{\sin\theta\, r^2}
\ea
with $\hat \ell = [\ell - qk (\cos \theta - 1)]/\sin\theta$.
Integrating these the following orbits are obtained
\be
 \varphi-\varphi_0 = \int \frac { dx}
 {\sqrt{-{\cal A} x^2 + {\cal B}x + {\cal C}} }\,.\label{orbita}
\ee
with ${\cal A} = (1 + k^2/\hat\ell^2)\sin^2\theta$, ${\cal B} =
(\veps^2 - u_i^2 - 2 ) k \sin^2\theta/\hat\ell^2$ and ${\cal C} =
(\veps^2 - u_i^2 - 1)\sin^2\theta/\hat\ell^2$.

We see that the motion is in the surface of a cone whose axis is
given by $\vec J$, with the source brane at the origin. The
trajectories can be obtained directly from eq.~(\ref{orbita}),
which, since the argument of the square root is quadratic in $x$,
can be easily integrated to give conic sections ({\it i.e.}
parabolic or hyperbolic motion) on the plane perpendicular to
$\vec J$.

\subsection{Equivalent Non-relativistic Lagrangians}
Since the orbits of these relativistic systems can be explicitly
expressed in terms of elementary integrals, it is possible to
identify a non-relativistic lagrangian of the form ${\cal L}_{\rm
eq} = \frac12 \dot{\vec y}^2 - V_{\rm eq}(\vec y)$ whose classical
trajectories precisely coincide with the trajectories of the fully
relativistic system. We call the lagrangian of this equivalent
system, the \emph{Equivalent Non-relativistic Lagrangian}. The
existence of this system is useful, since general results often
exist for these non-relativistic systems which are derived under
the assumption of trivial kinetic energies. Although these cannot
be directly applied to the relativistic $D$-brane systems, due to
the non-trivial velocity-dependent interactions which they
contain, their implications \emph{can} be applied to the
equivalent non-relativistic system, and so carried over to the
trajectories of the fully relativistic system.

\ssubsubsection{Born-Infeld Only}
In the Born-Infeld case, the (fully relativistic) trajectories are
given by eq.~\pref{orbit}. These same trajectories also give the
orbits for the system defined by the non-relativistic Lagrangian
having the following form
\be
    {\cal L}_{\rm eq}(r)
    =\frac 12 (\dot r^2 + r^2 \dot \varphi^2) +
    \frac{k_{p-p'}}{r^{7-p}}\,,
\ee
for which $V_{\rm eq} = - k_{p-p'}/r^{7-p}$. The orbits of this
system agree with the relativistic one provided we make the
identifications $k_2 = (\varepsilon^2-u_i^2) k/2$, $k_4 =
(\varepsilon^2-u_i^2-1) k/2$ and identify the conserved energy as
$\varepsilon_{\rm eq} = (\varepsilon^2-u_i^2-1)/2$.

As mentioned above, the utility of identifying this equivalent
system comes from the general results which are available in the
non-relativistic case. For instance \cite{goldstein}, general
central force problems, $V_{\rm eq} \propto r^\xi$ are known to
have solutions in terms of trigonometric functions when the
exponent is $\xi = 2,-1$ or $-2$. This corresponds in the brane
case to $p = 5$ or $6$ since $\xi = p-7$. Alternatively, the
solution may be written in terms of elliptic functions when $\xi =
6,4,1,-3,-4$ or $-6$ (corresponding to $p=1,3$ and $4$).
Furthermore, the general result known as Bertrand's theorem states
that the only potentials giving rise to closed orbits are the
Kepler problem, $\xi = -1$ (or $p = 6$), and the harmonic
oscillator, $\xi = 2$ (which would be $p=8$, and so has no
analogue within the domain of our approximations). This allows us
to conclude that only for $p=6$ can branes have closed orbits, and
then only when the energy is negative, $\varepsilon_{eq} < 0$. Looking
at eq.(\ref{Hamil}), we see that, as expected,  this last condition can be 
satisfied only if $p'=4$ and is never realized
in the BPS case $p'=2$. Being the orbits non precessing,  
we know that we must have some conserved
vector analogous to the Runge-Lenz vector 
of the nonrelativistic system, this vector can be found 
in a straightforward manner following 
similar steps as for the branonium case \cite{branonium1}.

Notice that the equivalent non-relativistic lagrangian has a
nontrivial potential even for the BPS case where $p - p' = 4$,
despite the vanishing of the potential in the relativistic
problem. The existence of the non-relativistic potential reflects
the fact that the trajectories of the non-relativistic system must
agree with the exact trajectories of the relativistic system, and
so the non-relativistic potential must encode the
velocity-dependent interactions of the moving $D$-brane system.
Notice that $k_4$ (and so also $V_{\rm eq}(r)$) does vanish for
\Dpp-branes which move only parallel to the \Dp-brane, since in
this case $\veps^2 = 1 + u_i^2$. In particular it also vanishes
for completely static branes, for which $u_i = 0$.

These observations allow a simple evaluation of the trajectories
of BPS branes moving only under the influence of
velocity-dependent forces. For example, as applied to the BPS
brane-brane system with $p = p'$ of ref.~\cite{branonium1}, the
trajectories of the fully-relativistic system are given by
elementary integrals, which reduce to conic sections (hyperbolae
and parabolae) when $p = 6$.

Remarkably, in the non BPS cases $p-p'=2$ the potential of the equivalent
non-relativistic Lagrangian has the same functional form as the
potential which arises in the non-relativistic limit of the full
Lagrangian. As a consequence of that, the functional form of the trajectories
do not change as one crosses over to the relativistic from the
non-relativistic regimes. Instead, the only change is how orbital
parameters (such as eccentricity and semi-major axis) depend on
physical quantities (like energy and angular momentum). Defining
the non-relativistic limit by $\veps \approx 1 + \veps_{NR}$, with
$\veps_{NR} \sim u^2_i \sim \ell^2/r^2 \sim k/r^{7-p} \ll 1$, we
see that in this limit $k_2 \approx k/2$, and so $V_{\rm eq}(r)$
smoothly goes over to $V_{NR}(r)$. This is as might have been
expected, given that the equivalent non-relativistic system must
incorporate in particular the non-relativistic trajectories of the
full $D$-brane problem. A similar correspondence was also found
for brane-antibrane motion with $p = p'$ \cite{branonium1}.

The same equivalence of functional form of $V_{\rm eq}(r)$ and the
non-relativistic potential, $V_{NR}(r)$, \emph{cannot} hold for
the BPS case, since we've just seen that $V_{\rm eq}(r)$ is
nonzero, while we know the non-relativistic limit of the BPS
problem has $V_{NR}(r) \equiv 0$. Instead in this case we only
have $V_{\rm eq}(r) \to V_{NR}(r) = 0$ in the non-relativistic
limit, since $k_4 \approx 0$.

\ssubsubsection{The Wess-Zumino term}
The (fully relativistic) orbits for the $D$6-$D$0 system including
the Wess-Zumino term are also the same as those obtained from an
equivalent non-relativistic problem, which in this case
corresponds to the Lagrangian of an electric charge in a field
generated by a dyon. The equivalent lagrangian is given by
\be
 {\cal L}_{\rm eq} = \frac 12(\dot r^2+r^2 \sin^2\theta_{\rm eq}\dot
 \varphi^2) + \frac {k_6}{r} -
 q\,k_6(\cos\theta_{\rm eq} -1)\dot \varphi
\ee
This problem has been
studied in detail in the past, see for instance \cite{boulware}.
The orbits obtained for this equivalent system are given in terms
of the conserved quantities $\ell_{\rm eq}$ and $\veps_{\rm eq}$
by
\be
 \varphi-\varphi_0 = \int \frac{dx}
 {\sqrt{-{\cal A}_{\rm eq} x^2 + {\cal B}_{\rm eq} x + {\cal C}_{\rm
 eq}}}
 \,,\label{orbitanr}
\ee
%
%
where ${\cal A}_{\rm eq} = \sin^2\theta_{\rm eq}$, ${\cal B}_{\rm
eq} = 2 k_6 \sin^2\theta_{\rm eq}/\hat\ell_{\rm eq}^2$, ${\cal
C}_{\rm eq} = 2 \veps_{\rm eq} \sin^2\theta_{\rm
eq}/\hat\ell^2_{\rm eq}$ and $\hat\ell_{\rm eq} = [\ell_{\rm eq} -
qk_6 (\cos\theta_{\rm eq} - 1)]/\sin\theta_{\rm eq}$. These integrals
coincide with the fully relativistic ones, (\ref{orbita}), if we
identify $k_6 = (\varepsilon^2-u_i^2-2)k/2$, 
$\veps_{\rm eq} = (\veps^2 - u_i^2 - 1)/2$,
$\hat\ell^2_{\rm eq} = \hat\ell^2 + k^2$ and 
$\sin^2\theta_{\rm eq} = \sin^2\theta(1
+k^2/\hat\ell^2)$. Nevertheless, it should be noted that this is not a
complete identification of the relativistic and the non-relativistic
orbits, but only of their projections into
the $(x,y)$ plane, because we are changing the angle of the cone in
which the motion takes place
$\theta \neq \theta_{eq}$.


\subsection{Quantum Dynamics}
Let us study the same problem from the quantum  mechanical point
of view. In the absense of a WZ term, the Hamiltonian will be
given by (\ref{Hamil}) and then the quantum mechanical wave
function satisfy
\be
 \sqrt{ h^{({p-{{p'}}}-4)/2} + u_i^2 + \frac 1 h
 \left( \rho^2 + \frac{\ell^2}{r^2} \right)}\Psi=
 \varepsilon \Psi \, .
\ee
Squaring the Hamiltonian operator we find
\be
 \left( h^{({p-{{p'}}}-4)/2} + u_i^2 + \frac 1 h
 \left( \rho^2 + \frac{\ell^2}{r^2} \right)\right)\Psi=
 \varepsilon^2 \Psi \, .
\ee
Taking into account that $p-p'$ can only be $2$ or $4$, we can
rewrite the above equation in the form
\be
\left(\frac12\left(\rho^2 + \frac{\ell^2}{r^2}\right)
-\frac {k_{p-p'}}{r^{7-p}} \right)\Psi=
\varepsilon_{eq}  \Psi \, .
\ee
where $\varepsilon_{eq}$ and $k_{p-p'}$ are defined as in the
previous section. Then, as in the $p=p'$ case, we recover the
equivalence of our fully relativistic problem with a
non-relativistic one at the quantum level.

A similar treatment can be done for the $p=6,\ p'=0$ case, in this
case the Schr\"odinger equation turns out to be:
\be
 \left(\frac{1}{2}\left(\rho^2+
  \frac{\hat\ell^2+k^2}{r^2}\right)
 -\frac{k(\varepsilon^2-\mu_i^2-2)}{2r}\right) \, \Psi\ =\
 \frac{\varepsilon^2-\mu_i^2-1}{2}\ \Psi
\ee
The equivalence with the corresponding non-relativistic system can
be read by making the same identifications as in the classical
case in the previous section. Notice that now, even though
there is a mapping to the corresponding equation of the
non-relativistic system, the mapping includes a shift in the
angular momentum $\hat\ell$. This is what prevents the complete
identification of the classical orbits, and it does not occur in
the $p-p'=2,4$ cases, the main technical reason for it is that in
this case there is a $h^2$ term is the square root in
(\ref{orbita}) that does not appear in the other cases. A similar
situation occurs in hydrogenoid atoms \cite{schweber}.

Notice that in all cases, this naive quantum treatment is limited
if, as we expect, the D-branes involved have macroscopic sizes.
The corresponding Bohr radius is inversely proportional to the
mass of the probe branes
%
%
%
and therefore this naturally sets us to the classical regime. The
only exceptions would be if the branes extend only over
compactified dimensions that could be either the compact
dimensions of compactified string theory, or the spacetime
dimensions in the very early universe. It would be interesting to
explore if in those cases this quantum behaviour has any physical
implications.

\section{Cheshire Branes: Uplifting to 11 Dimensions}
It is well known that Type IIA strings may be obtained by
compactifying M-Theory on a circle, but there are no branes in M
theory to which $D6$-branes can lift. They do not lift to 11
dimensions as branes at all, but instead as particular  multi
Taub-NUT spaces. We break from the previous line of development in
this section to use this lift to provide a complementary analysis
of branonium-like systems in M-theory.

\subsection{The Taub-NUT Background and Probes}
The first step is to define the 11-dimensional field configuration
which corresponds to the uplift of a stack of source $D6$-branes
(together with the fields to which they give rise).

\ssubsubsection{The background}
The eleven dimensional background obtained when uplifting $N$
$D6$-branes is purely gravitational, since the ten-dimensional
dilaton, Ramond-Ramond 1-form and the metric are encoded in the 11
dimensional metric as follows:
\be
    ds_{(11)}^2 = ds_{Mink7}^2 + h\, d\vec x \cdot d\vec x
    + h^{-1}(d\psi+\vec
    A \cdot d\vec x)^2
\ee
Here $ds_{Mink7}^2$ represents seven-dimensional flat space, and
the remaining part of the metric is a four-dimensional Euclidean
Taub-NUT metric. The $\psi$ coordinate is the 11th dimension, and
is a periodic variable with period $4\pi  l_s$, while $h$ and
$\vec A$ are given by
\be
    h = 1+ \frac{k}{r} \,, \qquad
   \vec \nabla\times \vec A = \vec \nabla h \,.
\ee
where  $k= g_s l_s N/2$.
The expression for $\vec A$ can be made more explicit by writing
instead $A_\varphi=  k  \, \cos\theta$.

%
%

\ssubsubsection{Probe Brane Actions}
The probes available in M-theory are $M2$-branes and $M5$-branes
which from the IIA point of view can describe $D2$-branes,
$F1$-strings, $NS5$-branes or $D4$-branes, depending on whether or
not these $M$-branes are transverse to or wrap the M-theory
dimension.

Let us consider a general Nambu-Goto type of action for the
M-theory branes in a purely geometric background:
\ba
    S &=& -M_{p'}\int d^{p'+1}\xi\,
    \sqrt{|g_{AB}(X)\,
    \partial_\alpha X^A\, \partial_\beta X^B|} \nn\\
    &&\qquad\qquad\qquad\ \qquad\qquad
    \alpha,\beta=0,...,p'\;;\qquad A,B=0,...,10 \,.
\ea
As in the previous section we can gauge-fix the reparameterization
invariance using static gauge and assume rigid motion, $X^a =
X^a(t)$, to get
\ba
    S &=& -m\int dt \, \sqrt{|g_{\alpha\beta}(X)+
    g_{a(\alpha}(X)\,\delta_{\beta)}^0 \dot X^a + g_{ab}(X)\dot X^a
    \dot X^b\, \delta_\beta^0\delta_\alpha^0|} \n &&
    \qquad\qquad\qquad\ \qquad\qquad \alpha,\beta=0,...,p'\;; \qquad
    a,b=p'+1,...,10 \label{accionn} \,.
\ea

There are two cases to consider, depending on whether the branes
are positioned at a single point in the $\psi$ dimension, or
whether they instead wrap this dimension.

\subsection{$D2$ and $NS5$ branes}
We now specialize to an M-theory brane which is located at a point
in the 11th dimension, and so can describe a $D2$-brane or an
$NS5$-brane as seen from the Type IIA point of view.

\ssubsubsection{The action}
In this section we focus on the case of a $(p'+1)$-dimensional
probe with $p'\leq 6$, and we assume that the $p'+1$ directions
along which it extends are transverse to the Taub-NUT space. The
choices $p' = 2$ and $p' = 5$ then describe the two choices for
M-theory branes. We have then $g_{\alpha\beta}(X) =
\eta_{\alpha\beta}$ and the probe-brane action becomes
\ba
S&=&-m \int dt\, \sqrt{1-\dot x_i^2- h(\dot r^2 +
  r^2(\dot\theta^2 +
    \sin^2\theta\,\dot\varphi^2))- h^{-1}(\dot\psi +
    k\cos\theta\,\dot\varphi )^2} \n&&
    \qquad\qquad\qquad\qquad\qquad\qquad\qquad\qquad
    \qquad\qquad\qquad
    i=p'+1,...,6 \,.
\ea
Using rotational invariance to place the plane of motion at
$\theta = \pi/2$, we have
\be
    S=-m\int dt\, \sqrt{1-\dot x_i^2- h(\dot r^2 +
    r^2\dot\varphi^2)-h^{-1}\dot\psi^2} \,.
    \label{accion}
\ee

\ssubsubsection{Equations of motion}
The canonical momenta derived from the action just given are
\ba
    p_i &=&  \frac {m} {\sqrt{(\cdots)}}\,  \dot x_i\,\,
    \equiv\,  m\, u_i
\n
    p_r &=&  \frac {m} {\sqrt{(\cdots)}}\, h \dot r\,\,
    \equiv\,  m\, \rho
\n
    p_\varphi &=&\frac {m} {\sqrt{(\cdots)}}\, h r^2 \dot\varphi\,\,
    \equiv\,  m\, \ell
    \n
    p_\psi &=&\frac {m} {\sqrt{(\cdots)}}\, h^{-1} \dot\psi\,\,
    \equiv\,  m\,\tilde \ell\,,
\ea
where now $\sqrt{(\cdots)}$ stands for the square root in
(\ref{accion}). In terms of these the energy becomes
\ba
    E &=&
    m{\sqrt{1+u_i^2+\left(\rho^2 + \frac 1 h \frac{\ell^2}{r^2}\right)
    + \tilde \ell^2 h}}
    \equiv m\, \varepsilon \, .
\ea

The equations of motion for this system then reduce to the
following first-order equations:
\ba
    \dot x_i &=& \frac {u_i} \varepsilon
    \n
    \dot r &=& \frac{\rho}{\veps h} = \frac 1{\varepsilon h}
    \sqrt{(\varepsilon^2-u_i^2 -1)h -\frac {\ell^2}{r^2}   -
    \tilde \ell^2 h^2 }
    \n
    \dot\varphi&=& \frac{\ell}{\veps hr^2}
    \n
    \dot \psi&=& \frac{\tilde \ell  h}{\varepsilon}
    \,.
\ea
These integrate to give the following trajectories:
\ba
    \varphi-\varphi_0&=&\int \frac {\ell dx} {\sqrt{
    (\varepsilon^2-u_i^2 -1)h  - \tilde \ell^2 h^2  -{\ell^2}{x^2}
    }}
    \n
    \psi-\psi_0 &=& \int   \frac {h^2 \tilde \ell dx} {x^2\sqrt{
    (\varepsilon^2-u_i^2 -1)h  - \tilde \ell^2 h^2  -{\ell^2}{x^2}
    }} \,.
\ea
Since $h$ is linear in $x = 1/r$, it is clear that the first of
these integrals gives conical sections. It is also easy to check
that there are no bound states, and so all orbits are parabolae or
hyperbolae as expected from the analysis of D6-D2 system in Type
IIA. Nor can Kaluza-Klein momenta change this conclusion.


\pagebreak
\ssubsubsection{Ten dimensional point of view}
How does the above picture compare with the same problem as
studied from a ten-dimensional point of view? As we have
previously seen, after gauge fixing and assuming rigid motion the
action for a $Dp'$-brane moving in a $D6$-brane background (with $p'
\neq 0$) is
\ba
    S &=& -m\int dt \,h^{(2-p')/4}
    \sqrt{1-\dot x_i^2-h (\dot r^2 + r^2 \dot \varphi^2 )}\,,
    \label{10dim}
\ea
which {\it differs} from the action (\ref{accion}) obtained from
the eleven dimensional point of view by the presence of the factor
$h^{(2-p')/4}$ outside the square root, and by the absence of the
last term, $h^{-1}\dot \psi^2$, inside the square root.

\medskip \noindent
{\it $D2$-Branes:} The first of these differences is absent if we
specialize to the $D2$-brane case, for which $p'=2$, as we now do.
To eliminate the second difference, we turn on an electromagnetic
field on the brane and rewrite the ten-dimensional $D2$-brane
action as follows \cite{Schmidhuber:1996fy}:
\be
    S = -T_2\int d^3\xi \left(e^{-\phi}\sqrt{|g_{\alpha\beta} +
    e^{2\phi}t_{\alpha}t_{\beta}|} + \frac 1 2
    \epsilon^{\alpha\beta\gamma}t_\alpha F_{\beta\gamma}\right)\,
\ee
which is equivalent to the usual BI form after eliminating
$t_\alpha$ using its equations of motion.

We now choose a purely magnetic field, $F_{12}=B$, and evaluate
using the $D6$-brane background (\ref{background}). In this case,
the eqs. of motion for $t_\alpha$ imply $t_1 = t_2=0$ and we have
\be
    S= -m\int dt \left( \sqrt{1-\dot x_i^2-h (\dot r^2 + r^2
    \dot \varphi^2  )-h^{-1}t_0^2  } + t_0 B \right) \,.
    \label{ho}
\ee
Integrating the last term by parts, the equations of motion for
$A_1$ and $A_2$ imply $\partial_1 t_0 = \partial_2 t_0 = 0$, and
so we define $t_0 = t_0(t) \equiv \dot\psi$. We obtain in this way
\be
    S= -m\int dt \sqrt{1-\dot x_i^2-h (\dot r^2 + r^2 \dot
    \varphi^2 )-h^{-1}\dot\psi^2  } \,,
\ee
which recovers the action obtained above from the purely
eleven-dimensional point of view.

The relation between the magnetic field and our eleven-dimensional
variables is obtained from the equation of motion for $t_0$ (now
called $\dot\psi$) derived from (\ref{ho}), which is
\be
    \tilde \ell=\frac{h^{-1}\dot\psi}{\sqrt{(\cdots)}} = B \,.
\ee
We see that the magnetic field on the $D2$-brane, as seen from the
ten-dimensional point of view, corresponds in the
eleven-dimensional picture to the momentum of the membrane in the
compactified direction. This is consistent with the interpretation
of the Kaluza-Klein momenta, $p \sim n/g_sl_s$, in the
compactified direction of M-theory as $n$ $D0$-branes. These
$D0$-branes form a bound state with the $D2$-branes that appears
as a magnetic flux on the $D2$-brane with first Chern number $c_1
= n$.

\medskip \noindent {\it NS5-Branes:}
We now return to the second possibility: the case of an $NS5$-brane
moving in a $D6$-brane background. In this case, following
\cite{Bandos:2000az}, we take the action to be
\ba
    S &=& - M\int d^4\xi \, e^{-2\phi}\sqrt{|g_{\alpha\beta}+ (\dot X^\mu
    \dot X_\mu
    -e^{2\phi}\dot y^2) \delta_\alpha^0\delta^0_\beta|}
    \n
    &=& -m\int dt\, \sqrt{1-\dot x_i^2-
    h(\dot r^2+r^2\dot\phi^2)+ h^{-1}\dot y^2} \,.
\ea
Here $y$ is the world-volume scalar field as defined in
\cite{Bandos:2000az}, and we see that the identification
$y=\psi-\psi_0$ leads to our original eleven-dimensional action,
now corresponding to an $M5$-brane in a Taub-NUT background.

\subsection{$F1$-Strings and $D4$-Branes}
The other possibility to consider is that the branes wrap the 11th
dimension to produce an $F1$-string or a $D4$-brane in Type IIA
string theory. We study these possibilities in this section.


\ssubsubsection{The action}
Taking the probe brane to be wrapped in the compact Taub-NUT
direction, $\psi$, and evaluating the determinant in
(\ref{accionn}) we obtain
\ba
    S &=& -m \int dt\, \sqrt{h^{-1}( 1-\dot x_i^2) - \dot r^2 -
    r^2 \dot \varphi^2} \label{ecua} \qquad \qquad i=p'+1,...,6 \,.
\ea


\ssubsubsection{Equations of motion}
Following similar steps as before, we obtain the canonical momenta
\ba
  u_i &=&  \frac {h^{-1} \dot x_i} {\sqrt{(\cdots)}}\, \,\n
       \rho &=&  \frac {\dot r} {\sqrt{(\cdots)}}\, \,\, \n
   \ell &=&\frac {r^2 \dot\varphi} {\sqrt{(\cdots)}} \,,
\ea
and hamiltonian
\be
    \varepsilon = \sqrt{u_i^2 + \frac 1 h \left(1+ \rho^2+
    \frac{\ell^2}{r^2}\right)}\,.
\ee
These lead to the first-order equations of motion
\ba
     \dot x_i &=& \frac {u_i}\varepsilon \n
    \dot r &=&  \frac 1 {\veps  h} \sqrt{(\varepsilon^2-u_i^2)h
    - 1 - \frac{\ell^2}{r^2}} \n
   \dot \varphi &=&  \frac {\ell} {\veps  h r^2}\,,
\ea
whose integration leads to the following trajectories
\ba
    \varphi-\varphi_0 &=&\int \frac {\ell \, dx}
    {\sqrt{(\varepsilon^2-u_i^2)h-1-\ell^2x^2} } \,.
\ea

\ssubsubsection{Ten dimensional point of view}
The comparison between 10 and 11 dimensions is in this case
simpler.

\medskip\noindent {\it $D4$-Branes:}
If we set $p=4$ in equation (\ref{10dim}) describing a $Dp$-brane
probe in a $D6$-brane background, we immediately obtain exactly
(\ref{ecua}). Our previous solution then represents, from the 10
dimensional point of view, a $D4$-brane having vanishing
world-volume gauge fields, moving in a $D6$-brane background.

\medskip \noindent {\it $F1$-Strings:}
On the other hand, if we write the Nambu-Goto action for a
fundamental string in a brane background, we obtain
\be
    S = -m\int dt\, \sqrt{|g_{\alpha\beta}(X)+
    g_{a(\alpha}(X)\,{\delta_{\beta)}}^0 \dot X^a + g_{ab}(X)\dot X^a
    \dot X^b\, \delta_\beta^0\delta_\alpha^0| } \,,
\ee
where, as before, we gauge fix using the static gauge and assume
rigid motion. Specialized to the $D6$-brane background, we also
immediately recover (\ref{ecua}), showing that our above solution
also represents a fundamental string moving in a $D6$-brane
background.

\section{Orientifold Atoms}
We next consider the motion of \Dp-branes in the presence of
orientifold planes. Orientifold planes are generic in type II string
compactifications and can interact with \Dp-branes.
An important difference in this case is the
potential breakdown of the description of a probe brane moving in
the fixed supergravity background due to the source branes, since
we typically only have a single orientifold plane as a source.

\subsection{Orientifold planes}
We consider a configuration of $N$ anti Dp-branes (as counted on
the covering space, {\it i.e.} $N/2$ dynamical branes) and an
orientifold $O^{\pm}$-plane. Each of these objects breaks one
different half of the supersymmetries of the bulk. The tension and
Ramond-Ramond charges of these objects are:
\begin{itemize}
\item anti-\Dp-branes: $T = 1$ and $Q = -1$,
\item $Op^{+}$-plane: $T = 2^{p-4}$ and $Q = 2^{p-4}$,
\item $Op^{-}$-plane: $T = -2^{p-4}$ and $Q = -2^{p-4}$.
\end{itemize}
{}From our analysis above we expect that at large distances the
antibranes are attracted to the $O^{+}$-plane (which has the same
mass and opposite charges) and are repelled by the $O^{-}$-plane.

The gauge group associated with $N$ antibranes plus an orientifold
plane is $USp(N)$ for the $O^{+}$-plane, and $SO(N)$ for the
$O^{-}$-plane. Scalars parametrizing the positions of the
antibranes transform in symmetric (antisymmetric) representations
if the gauge group is $USp(N)$ ($SO(N)$). The fermions transform
the other way around, being in the antisymmetric representation
for the $O^{+}$-plane and in the symmetric representation for
$O^{-}$-plane.

\ssubsubsection{The M\"obius strip amplitude}
Let us first consider the case where the branes and orientifolds
are attracted to one another: $Op^+$ planes and $Dp$-antibranes.
We compute their interaction by evaluating the disk and cross-cap
amplitudes given the tension and Ramond-Ramond charges of the
antibranes and orientifold planes respectively. The contribution
to the vacuum energy of each of these amplitudes is:
\begin{equation}
{\cal D} = \bar{N} T_{Dp} V_p
\end{equation}
\begin{equation}
{\cal C} = 2^{p-4} T_{Dp} V_p \,,
\end{equation}
where $\ol{N}$ is the number of anti-\Dp-branes, $T_{Dp}= [{g_s (2
\pi)^p l_s^{p+1}}]^{-1}$ is the tension of a \Dp-brane and $V_p$
is the volume of the dimensions parallel to our objects (for
convenience we consider these dimensions to be toroidally
compactified). Both objects have positive tension so the total
tension at this order is,
%
%
\begin{equation}
{\cal C} + {\cal D} =  (\bar{N} + 2^{p-4}) T_{Dp} V_p
\end{equation}

The next order correction (one loop in open string counting) comes
from the cylinder, the M\"obius strip and the Klein bottle
\cite{oldies}. The cylinder and the Klein bottle amplitudes
are zero, because a nonzero result requires both a boundary and a
cross-cap amplitude since it requires the minimum combination
which breaks all of the bulk supersymmetry. This leaves the
M\"obius strip as the only nonzero amplitude:
\begin{equation}
    {\cal M} = - \frac{\ol{N} V_p}{l_s^{p+1}} \int_0^{\infty}
    \frac{dt}{t} \, \frac{e^{{-2 r^2
    t}/({\pi l_s^2})} }{2 (8 \pi t)^{({p+1})/{2}}}  \; F(q^2) \,,
\end{equation}
where $r$ is the separation between the antibrane and the
orientifold plane and $F(q^2)$ is the contribution from the string
oscillator modes:
%
%
\begin{equation}
F(q^2) = \left( \frac{f_2(q^2)f_4(q^2) }{f_1(q^2) f_3(q^2)}
\right)^8 = \frac{\theta_{10}^4(0|-it) \theta_{01}^4(0|-it)}{
\eta^{12}(-it) \theta_{00}^4(0|-it)} \,.
\end{equation}
Here $q = e^{- \pi t}$ and $F$ has the asymptotics $\lim_{t
\rightarrow \infty} F(q^2) = 16$ and  $\lim_{t \rightarrow 0}
F(q^2) = 256 t^4$.

Large separations between the branes and orientifold plane
corresponds to the UV in the open string picture and IR in the
closed string picture. In this limit the main contribution arises
from the exchange of massless closed-string modes: the graviton,
dilaton and Ramond-Ramond fields. This corresponds to the limit $t
\rightarrow 0$ in the integral (large t contribution are
suppressed in the integral). The contribution of the oscillator
modes reduces to $F(q^2)\rightarrow 2^{8} t^4$
and the integral becomes
\begin{equation}
    {\cal M} = -   \frac{2^8 \ol{N} \, V_p}{l_s^{p+1}} \int_0^{\infty}
    \frac{dt}{t} \, \frac{t^4 \, e^{{-2
    r^2 t}/({\pi l_s^2})} }{2 (8 \pi t)^{({p+1})/{2}}}
    = - 2^{p-4}\ol{N} g_s T_p V_p \left[ 2^8 \pi
    \Gamma \left( \frac{7-p}{2} \right) \right]
    \left( \frac{l_s}{r} \right)^{7-p} \,.
\end{equation}

This is the expected result: the power-law associated with the
exchange of massless closed-string states. In this limit the
result can be obtained by the supergravity approximation in the
weak-field approximation, and is the same as is obtained in the
branonium case.

The difference with the branonium case comes at short distance.
For branonium an open string tachyon appears at distances of the
order of the string scale, signaling the decay of the antibrane
with one of the branes in the stack of source branes. In the
orientifold case there is no tachyon at short distance, so the
system is expected to be stable. At short distance the best
description is found in the open-string picture, and at the
massless level we find a field theory description in $p+1$
dimensions consisting of a $USp(\ol{N})$ gauge group with $9-p$
scalars in the adjoint representation ({\it i.e.} the fields
parameterizing the position of the branes) and $2^{9-p}$ fermions
of both chiralities (in even dimensions).

To obtain the potential for the adjoint scalars in this regime we
must go to the $t \rightarrow \infty$ limit. There
$F(q^2)\rightarrow 2^{4}$ and the potential takes the form:
\begin{equation}
    {\cal M} =  g_s  V_p T_p \sum_{n=0} a_n \left( \frac{r}{l_s}
    \right)^{2n} \,,
\end{equation}
where
\begin{equation}
a_n = (-1)^{n+1} \ol{N} k_{n - (p+3)/2} \, \left(
\frac{2^{n-(p+5)/2} \pi^{(p-n-1)/2}}{n!} \right) \,,
\end{equation}
and
\begin{equation}
k_j = \int dt \; t^j F(q^2) \,,
\end{equation}
is a positive number if $-5 < j < -1$. Numerically, $k_{-9/2} =
21.75$, $k_{-7/2} = 13.47$, $k_{-5/2} = 13.92$, $k_{-4} = 13.82$,
$k_{-3} = 12.31$.

{}From these expressions there are two results which bear
emphasis. First, the $a_0$ term describes a correction to the
vacuum energy, which is finite and negative for $p < 7$. We can
see that the one-loop correction is therefore acts to reduce the
tree-level tension. Second, the $a_1$ term gives a
harmonic-oscillator, $r^2$, potential near $r = 0$ and so
represents a positive mass for the adjoint fields representing the
\Dp-brane center-of-mass motion.

\subsection{Interaction Potential}
The potential which results from these considerations is drawn in
figure \ref{potential-orientifold}. A useful interpolation which
captures both the $1/r^{7-p}$ behaviour at long distances and the
$r^2$ behaviour at short distance is the Lorentzian
\begin{equation}
    V(r) = -  \, \frac{g_s  T_p \, |a_0|}{(1+ \alpha \,
    r^2)^{({7-p})/{2}}}
\end{equation}
where $\alpha$ is a constant that can be adjusted to match the
long-distance behaviour.

\FIGURE{
\centering
\hspace*{0in}
\vspace*{.2in}
\epsfxsize=3in
\epsffile{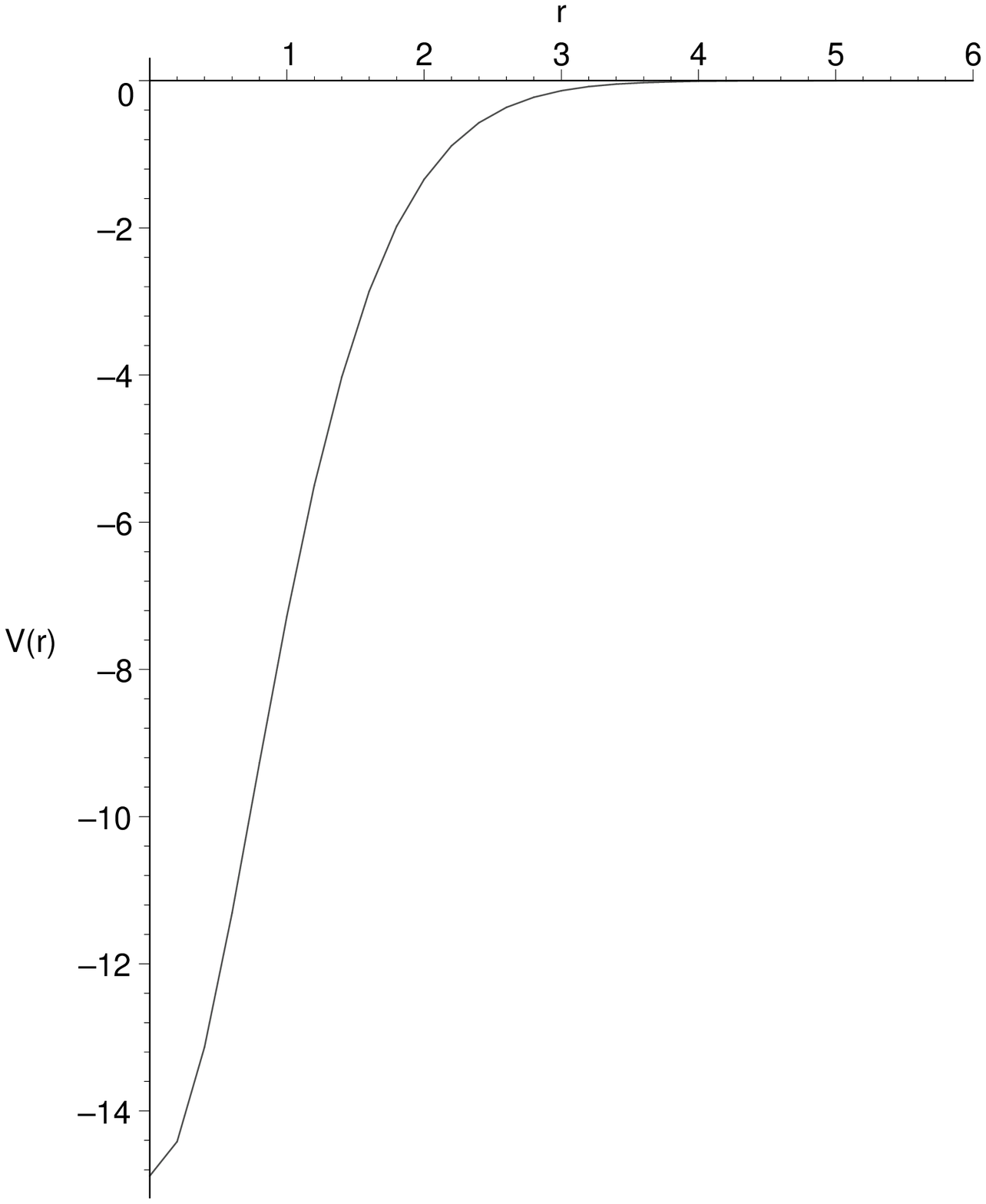}
\hspace*{0.4in}
\caption{\small Attractive potential for an orientifold plane
$Op^+$ and an antiD-brane from the M\"obius strip. At long
distances the result coincides with a brane-antibrane system
(potential of the form $1/r^n$). At short distances the system is
better understood in the open string picture where the scalar
fields representing the position along the transverse dimensions
take a mass.}
\label{potential-orientifold}
}

\pagebreak

\ssubsubsection{Long-distance regime}
The analysis in this regime follows the same lines as for
branonium. This is because, from the supergravity point of view,
the metric, dilaton and Ramond-Ramond forms cannot distinguish
between an orientifold plane and a set of $2^{p-4}$ branes. Notice
however that the limit of a large number of source $D$-branes
(which is possible for branonium) cannot be reproduced here. In
particular, for $p=5$ the orientifold has the same tension and
charge as a $D$-brane (taking into account the orientifold image).
Because of this we cannot access the regime where nonlinear
supergravity effects are important without leaving the regime of
validity of our approximations.

\ssubsubsection{Short-distance regime}
As explained earlier, the one-loop contribution gives a
short-distance potential of the form $a_1 r^2$, with $a_1 \sim
g_s/l_s^2$.

The one-loop vacuum amplitude is negative and so reduces the tree
level tension, giving
\begin{equation}
\Lambda =   [ (\bar{N} + 2^{p-4}) + g_s a_0] T_{Dp} V_p \,.
\end{equation}
We can see the limits of validity of the loop expansion by asking
for the value of the string coupling for which the tree-level and
one-loop contributions cancel is given by
\begin{equation}
g_s =   \frac{(\bar{N} + 2^{p-4})}{|a_0|} \,.
\end{equation}
Since $a_0 \sim - \ol{N}$, the transition between the two regimes
occurs at $g_s \sim 1$.

\subsection{Strong coupling regime}
Ref.~\cite{u} makes several conjectures concerning the behaviour
of this type of system at strong coupling. Let us consider the
case of a $O3^+$-plane and an antibrane (2 counted in the covering
space). The total Ramond-Ramond charge of the system is $q_{RR} =
1/2 -2 = -3/2$. The gauge group at low energies is $USp(2)=
SU(2)$, with some fermions which are singlets (because they
transform in the antisymmetric representation). There are also 6
scalars parameterizing the positions of the antibranes, which
transform in the adjoint representation.

This system is conjectured to be S-dual (the coupling in the
S-dual system is $\tilde{g_s} = 1/g_s$, and the string length is
$\tilde{l_s}= g_s^{1/2} l_s$ to keep the Plank mass fixed) to an
$O3^-$-plane containing one embedded anti-$D3$-brane. The
Ramond-Ramond charge of this system is $q_{RR} = -1/2 -1 = -3/2$.
The world-volume theory has no gauge bosons and scalars (recall
the antibrane is stuck at the position of the orientifold plane),
but there is a fermionic zero mode. As is explained in \cite{u}
this conjectured duality illustrates two mechanisms of keeping
$D$-branes close to orientifold planes: either by loop effects
through the correction to the masses of the center-of-mass
scalars, or by projecting out by the orientifold.

In the limit of very strong coupling the vacuum energy comes from
the tension of the objects with $T \sim 1/\tilde{g_s}$,
\begin{equation}
    \Lambda(g_s \rightarrow \infty) =   [-1/2 + 1] g_s \, \frac{V_p}{(2
    \pi)^{3} l_s^{4} g_s^2} \,.
\end{equation}
The first-order correction to this result comes from the M\"obius
strip. This gives a constant contribution (the tree level scalars
are projected out) that is the same as the $a_0$ of the S-dual
case, but with opposite sign. So the sum of the tree-level and
one-loop results at strong coupling is
\begin{equation}
    \Lambda(g_s \rightarrow \infty) =   ([-1/2 + 1] g_s + |a_0|)
    \, \frac{V_p}{(2 \pi)^{3} l_s^{4} g_s^2} \,.
\end{equation}
By contrast, at small coupling the vacuum energy goes like
\begin{equation}
    \Lambda(g_s \rightarrow 0) =   \left(
    [1/2 + 2] \frac{1}{g_s} - |a_0| \right)
    \frac{V_p}{(2 \pi)^{3} l_s^{4}} \,.
\end{equation}

In order to obtain the potential for the dilaton in the Einstein
frame we should multiply these results by $g_s$, so at weak
coupling we find
\begin{equation}
    \Lambda(g_s \rightarrow 0) =  ([1/2 + 2]  - |a_0| g_s)
    \, \frac{V_p}{(2 \pi)^{3} l_s^{4}} \,,
\end{equation}
while at strong coupling we have
\begin{equation}
    \Lambda(g_s \rightarrow \infty) =   ([-1/2 + 1] g_s + |a_0|)
    \,\frac{V_p}{(2 \pi)^{3} l_s^{4} g_s} \,.
\end{equation}

\begin{figure}[h]
  \def\epsfsize#1#2{0.8#1}
  \begin{center}
  \epsffile{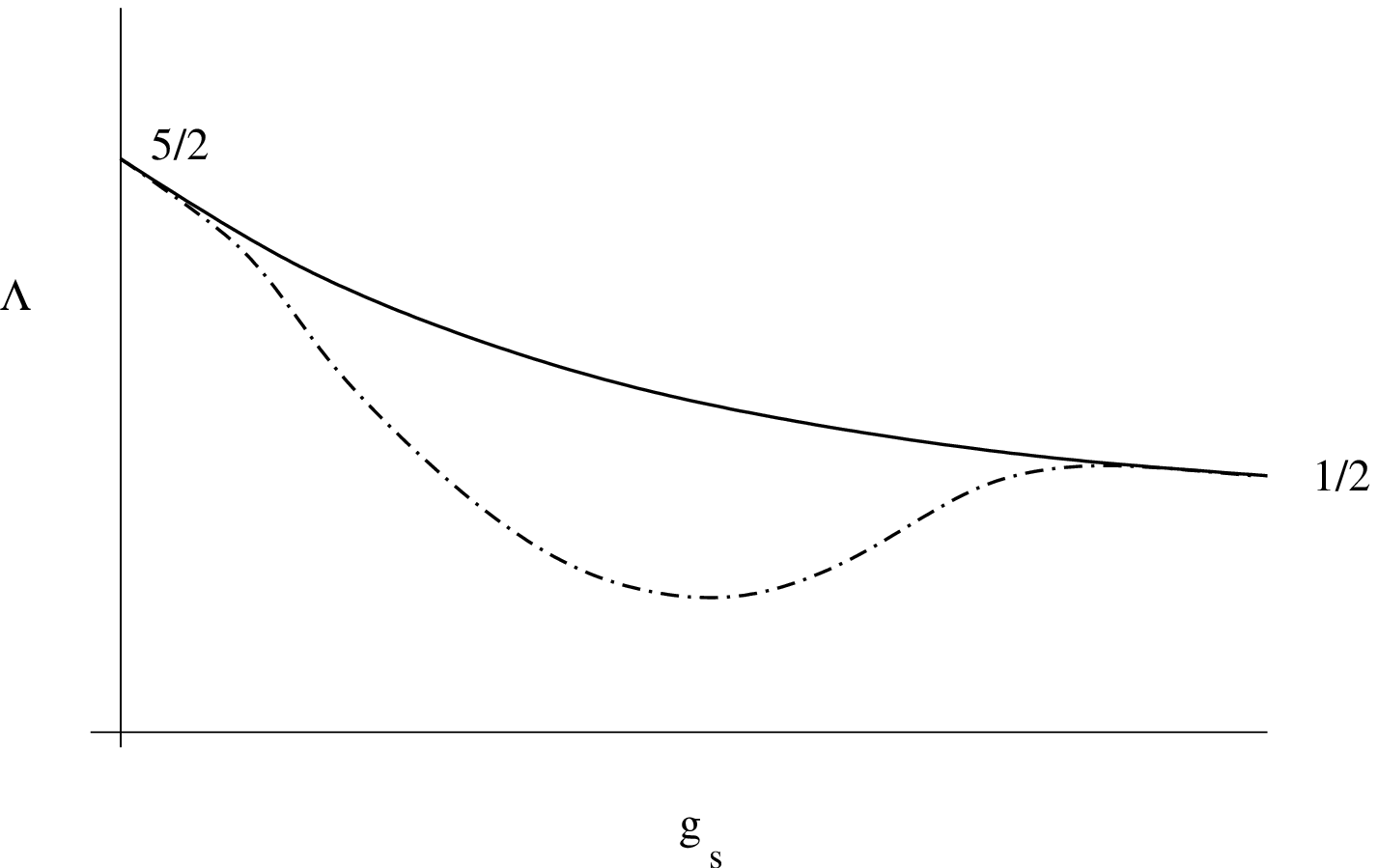} \label{vacuum}
  \caption{\it Schematic
representation of the vacuum energy for the $O3^+$-antibrane
system depending on the coupling. The two lines represent two
different possibilities: in the solid line case the system is
driven to stri=ong coupling while in the dashed line case the
dilaton is stabilized.}
\end{center}
\end{figure}


These results are what one might expect. If loop effects are
ignored there is no potential for the dilaton, because it is the
loop effects which know about the breaking of supersymmetry.
Consequently the potential is constant. The fact that the vacuum
energy at strong and weak coupling are not equal to one another
also has to do with supersymmetry breaking. If we take a
supersymmetric combination of $D3$-branes and orientifold 3-planes
the potential is the same and flat.

We plot in figure 6 the schematic shape of the vacuum
energy as a function of the coupling. It is not clear if there is
a value for the coupling for which the vacuum energy can be zero
or negative. It is also unclear if the dilaton is stabilized at
any finite value ({\it i.e.} as would be the case for the dashed
line in the figure) or not (in which case the $D$-brane is driven
onto the $O3^-$ plane system). Notice that once $g_s > 5/(2|a_0|)$
the weak-coupling expression for $\Lambda$ falls below the
asymptotic value of the strong-coupling expression at infinity,
and this would give evidence for the existence of a minimum to the
extent that this coupling is still within the domain of
applicability of leading-order perturbation theory.

Another interesting observation is that taking T-dualities of the
Sugimoto model \cite{s} and moving branes we can obtain a system
having 16 $O3^+$-planes with 2 anti $D3$-branes attached to them,
plus 48 $O3^+$-planes without branes. The total tension of this
system is $64/g_s$ in $D3$-brane units. By applying the S-duality
rules above one can get a system without dynamical D-branes,
consisting of 16 $O3^-$-planes with 1 anti-D3-brane (that cannot
move from it) and 48 $O3^-$-planes with a non-dynamical $D$-brane
on top of each one. The total tension of this dual system is $32
g_s$. The same considerations regarding the stabilization of the
coupling that were made in the non-compact case also apply here.

\subsection{Repulsive case}
We now consider the case with an orientifold plane $Op^-$ and an
antibrane. The arguments proceed exactly as in the previous
example, with the following differences:

\begin{itemize}
\item The open-string description implies an $SO(N)$ gauge group
and scalars in the adjoint (with fermions in symmetric)
representations,
\item The tension of the orientifold has an opposite sign (negative
tension), while the antibranes carry positive tension. This means
that the disk plus cross-cap amplitudes in this case are
\begin{equation}
{\cal C} =  (\bar{N} - 2^{p-4}) T_{Dp} V_p.
\end{equation}

\item The M\"obius amplitude gives a repulsive potential (the same
shape as before, but with a global minus sign). The antibranes are
then unstable if placed on top of the orientifold and the system
becomes unstable at one loop.
\end{itemize}

\section{Conclusions}
With this paper we begin a discussion of the motion of various
kinds of probe \Dpp-(anti)branes about a stack of \Dp\ source
branes  where $p' < p$, or orientifold planes. This study is a
natural extension of the branonium (\Dp-antibrane orbiting a stack
of \Dp-branes) analysis, of ref.~\cite{branonium1}, and the
remarkable existence we find here of simple integrable
relativistic motion generalizes the same property which was found
there. We call the bound states, when these exist, `branic' or
`oriental' atoms (depending on whether it is a source brane or
orientifold plane about which the orbits occur) to underline that
unlike branonium these systems are stable against mutual
annihilation.


For the \Dpp-\Dp-brane systems we examine the relativistic
equations of motion, and provide a solution by quadratures for the
fully relativistic orbits. The long-range force experienced by the
probe brane is attractive when $p - p' < 4$, is flat in the BPS
case where $p - p' = 4$ and is repulsive when $p - p' > 4$. These
three conditions also coincide with the conditions for where the
lowest level of the open-string sector of the theory is tachyonic,
massless or massive respectively. These trajectories can involve
bound states having Keplerian conic sections as orbits if $p=6$
and $p'=4,2$.

We examine the case of probe branes moving around $D6$ source
branes in a bit more detail, and in particular consider the
perspective which obtains if these systems are lifted to M-theory
in 11 dimensions. There are no $D6$-branes in M-theory, but the
fields they source in the IIA theory survive (like the Cheshire
Cat's smile) into 11 dimensions as a Taub-NUT-type metric
configuration, and we show explicitly how the orbits of probe
branes in this Taub-NUT space duplicate the motion of probes about
a $D6$ brane in IIA supergravity.

Finally, we consider some aspects of the interaction potential
experienced by an anti-brane/orientifold plane system, explicitly
exhibiting the cross-over from the long-distance, weak-field
regime to the short-distance, harmonic-oscillator regime.

It is interesting to speculate about the possible physical
implications of these systems, especially for early universe
cosmology. Understanding the interactions and relative motion of
$D$-branes may play an important role in scenarios of brane
dynamics in the early universe, such as brane gases \cite{BV},
$D$-brane inflation \cite{braneinflation} or alternatives
\cite{Ekpyrosis}, mirage cosmology \cite{mirage}, {\it etc}. For
this a more complete analysis of, at least, the most interesting
systems that we studied here (such as $D6$-$D2$ and
$D$-brane-orientifold) might be needed.

Some of the analyses done for the branonium case can in principle
also be generalised to these systems, such as to investigate
stability against perturbations in the transverse direction of the
probe brane, radiation into bulk and brane degrees of freedom,
orbital decay time, and so on. In particular, an understanding of
the behaviour of these systems after compactification is very
likely required for real applications.

It is striking that the $p-p'=2,4$ cases have the same classical
trajectories as equivalent non-relativistic systems, even though
the original systems are fully relativistic. A similar situation
also arose in the branonium system. For BPS systems the potential
of the equivalent non-relativistic lagrangian encodes the
velocity-dependent forces of the original relativistic problem.
For non-BPS systems the equivalent non-relativistic potential has
the same functional form as the full system's potential in the
non-relativistic limit, although this not true for the BPS case.
It is not clear why these systems are integrable, and whether it
is only the BPS cases for which the equivalent and
non-relativistic potentials differ. In the presence of the
Wess-Zumino term --- {\it i.e.} when $p=6$ and $p'=0$ --- the
equivalent system was the lagrangian for a charge moving in the
field of a dyonic. A further understanding of these relations is
certainly needed.


Finally we may consider yet more complicated systems. The relative
motion of an anti-brane in a lattice of \Dp-branes (which one
might call the `brane transistor') could have interesting
properties and possible applications in cosmology. Also systems of
intersecting branes, which are known to have interesting
phenomenological and cosmological properties, have similar
background geometries as those studied here and the discussion
presented here may be generalizable to these cases. There is no
shortage of directions deserving closer scrutiny.

\section{Acknowledgements}
We thank D. Mateos, R. Myers for interesting conversations. C.P.B.
is partially funded by NSERC (Canada), FCAR (Qu\'ebec) and McGill
University. F.Q. is partially funded by PPARC. N.E.G. is supported
by Fundaci{\'o}n Antorchas and CONICET, Argentina. C.B. and F.Q.
thank the Kavli Institute for Theoretical Physics in Santa Barbara
for their hospitality while part of this work was done. (As such
this work was partially supported by the NSF under grant number
PHY-99-07949.)

\end{document}